\documentstyle[12pt,epsfig,amsfonts,latexsym]{article}

\makeatletter

\@addtoreset{equation}{section} \makeatother

\newcommand{\be}{\begin{equation}}
\newcommand{\ee}{\end{equation}}
\newcommand{\ba}{\begin{eqnarray}}
\newcommand{\ea}{\end{eqnarray}}
\newcommand{\bea}{\begin{eqnarray}}
\newcommand{\eea}{\end{eqnarray}}
\newcommand{\bean}{\begin{eqnarray*}}
\newcommand{\eean}{\end{eqnarray*}}

\newcommand{\bk}{{\bf k}}
\newcommand{\bx}{{\bf x}}

\def\spose#1{\hbox to 0pt{#1\hss}}
\def\lsim{\mathrel{\spose{\lower 3pt\hbox{$\mathchar"218$}}
\raise 2.0pt\hbox{$\mathchar"13C$}}}
\def\gsim{\mathrel{\spose{\lower 3pt\hbox{$\mathchar"218$}}
\raise 2.0pt\hbox{$\mathchar"13E$}}}

\begin{document}



\begin{flushright}
LPT-ORSAY 03-53
\end{flushright}
\vskip 1cm
\begin{center}
{\Large \bf Testing and comparing tachyon inflation to single 
standard field inflation}
\end{center}
\vskip 1cm
\begin{center}
D.A.Steer $^{a}$\footnote{{\tt
steer@th.u-psud.fr}} and F.Vernizzi $^b$\footnote{{\tt f.vernizzi@iap.fr}}
\\
\vskip 5pt \vskip 3pt
{\it a}) Laboratoire
de Physique
Th\'eorique\footnote{Unit\'e Mixte de Recherche du CNRS (UMR 8627).}, B\^at. 210, Universit\'e
Paris XI, \\ 91405 Orsay Cedex, France\\
and
\\
F\'ed\'eration de recherche APC, Universit\'e Paris VII,\\
2 place Jussieu - 75251 Paris Cedex 05, France.
\vskip 3pt
{\it b}) Institut d'Astrophysique de
Paris, GR$\varepsilon$CO, FRE 2435-CNRS, 98bis boulevard Arago,
75014 Paris, France
\end{center}
\vskip 2.0cm

\setcounter{footnote}{0} \typeout{--- Main Text Start ---}



\begin{abstract}
We study the evolution of perturbations during the domination and
decay of a massive particle species whose mass and decay rate are
allowed to depend on the expectation value of a light scalar
field. We specialize to the case where the light field is
slow-rolling, showing that during a phase of inhomogeneous
mass-domination and decay the isocurvature perturbation of the
light field is converted into a curvature perturbation with an
efficiency which is nine times larger than when the mass is fixed.
We derive a condition on the annihilation and decay rates for the
domination of the massive particles and we show that standard
model particles cannot dominate the universe before
nucleosynthesis. We also compare this mechanism with the curvaton
model. Finally, observational signatures are discussed. A CDM
isocurvature mode can be generated if the dark matter is produced
out of equilibrium by both the inflaton and the massive particle
species decay. Non-Gaussianities are present: they are chi-square
deviations. However, they might be too small to be
observable.

\end{abstract}


\date{\today}

\renewcommand{\thefootnote}{\arabic{footnote}} \setcounter{footnote}{0}

\section{Introduction}
The recent WMAP data\cite{WMAP1} strongly supports the idea
that the early universe underwent a phase of inflation.
One typically considers an inflationary phase driven by the
potential or vacuum energy of a scalar field, the inflaton, whose
dynamics is determined by the Klein-Gordon action. 
More recently, however, motivated by string
theory, other non-standard scalar field actions have been used in
cosmology.
One particular type of field
which has attracted 
attention is the tachyon\cite{Gibbons1} $T$,
whose action is of the Dirac-Born-Infeld form\cite{action},
\be
S_T = - \int d^4 x \sqrt{-g}  V(T)  \left( 1 +
g^{\mu \nu} \partial_{\mu} T
\partial_{\nu} T  \right)^{1/2}, \quad {\rm sign.} \left\{ g \right\} =(-,+,+,+), 
\label{eq:action1}
\ee
where 
$V(T)$ is its potential.
According to Sen's conjecture, 
in type II string theory the tachyon signals the
instability of unstable and uncharged D-branes of tension
$\lambda=V(0)$.
In this context the positive potential
$V(T)$ is even and satisfy the properties
$ V'(T>0)<0, \qquad V(|T| \rightarrow \infty)
\rightarrow 0$.

Here (see also \cite{DF}) we take a phenomenological approach and study
the inflationary predictions of a phase of inflation driven by a
field $T$ satisfying the action Eq.~(\ref{eq:action1}) and we assume 
that the potential satisfies the properties mentioned above. We call
this tachyon inflation although the potential $V(T)$ may not be
particularly string inspired.  
The questions we address here are: 1) Does tachyon inflation lead
to the same predictions as standard single field inflation (SSFI)?
2) Can tachyon inflation already be ruled out by current
observations? 3) Can we discriminate between tachyon inflation and
SSFI in the light of new and planned future experiments?
The answer to the first question is no: tachyon inflation leads to
a deviation in one of the second order consistency relations.
However, tachyon
inflation cannot be ruled out at the moment, and its predictions
are typically characteristic of small field or chaotic inflation.
The answer to the final question is given in the last section.

\section{Slow-roll predictions of tachyon inflation}

In a homogeneous and isotropic
background with scale factor $a$,
the tachyon field can be
treated as a fluid with energy density and pressure 
$\rho  = {V(T)}/{(1 - \dot T^2)^{1/2}}$, 
$P = w \rho = - (1 - \dot T^2) \rho $. 
For tachyon inflation, the basic condition for accelerated
expansion is that
 \be \frac{\ddot a }{a} = -
\frac{1}{6 M_{\rm Pl}^2}(\rho + 3 P) =  \frac{1}{3 M_{\rm Pl}^2}
\frac{V}{(1-\dot{T}^2)^{1/2}} \left(1 -\frac{3}{2} \dot T^2
\right)
>0 \ \ \Rightarrow \ \ \dot T^2 < \frac{2}{3}.
\ee
In order to study inflationary predictions we must define slow-roll parameters.
Here we use the 
horizon-flow parameters\cite{Dominique} based on
derivatives of $H$ w.r.t.\ the number of $e$-folds $N=\ln a$.
This definition has the advantage 
of being independent of the field driving
inflation and thus it is a natural choice to use in order to compare
SSFI and tachyon inflation.  
We just need the first three parameters, which are
\be \epsilon_1   =  ({3}/{2}) \dot T^2
, \label{eq:eps} \quad 
\epsilon_2   = \sqrt{{2}/({3 \epsilon_1})} {\epsilon_1'}/{H}
=  2 {\ddot T}/({H \dot T}) , \quad
\epsilon_2 \epsilon_3 = \sqrt{{2 \epsilon_1}/{3 }} {\epsilon_2'}/{H},
\ee
where ${}'$ is the derivative w.r.t.\ $T$.
Thus as in SSFI, tachyon inflation is based upon the slow
evolution of $T$ in its potential $V(T)$, with the slow-roll
conditions
$\ddot T \ll  3H \dot T$ and $\dot T^2 \ll 1$.

It is well known that during an inflationary phase, quantum vacuum
fluctuations are stretched on scales larger than the horizon.
There, they are frozen until they reenter the horizon after
inflation. 
Calculation of the spectra of scalar quantum fluctuations proceeds
by defining a canonical variable which can be quantized with the
standard methods. The straightforward generalization of the
canonical variable to the case of a tachyon fluid is\cite{GarriMuka} 
$v_\bk=z M_{\rm Pl} (\psi + H \frac{\delta T}{\dot T})$,
where $\psi$ is the Bardeen potential in conformal gauge and 
$\delta T(t,\bx)$
is a linear perturbation around the homogeneous solution $T(t)$.
The pump field $z$ is defined by $
z = {\sqrt{3} a \dot T}/{(1 - \dot T^2)^{1/2}}
= -{a \sqrt{2 \epsilon_1}}/{w}$.
The equation derived from minimizing the action expanded to second
order in $v_\bk$ is\cite{GarriMuka}
\be
\frac{d^2{v}_\bk}{d\tau^2} -\left( w k^2 + U(\tau) \right)
{v}_\bk = 0, \ \ \ \ U(\tau) \equiv \frac{1}{z} \frac{d^2 z}{d
\tau^2}. \label{eq:canon}
\ee
Note the factor of $-w$, the speed of sound of
the tachyon fluid, in front of $k^2$, absent in SSFI where the speed of 
sound is 1. 
Instead of computing directly
$U(\tau)$ in terms of the slow-roll parameters, we
observe that in SSFI inflation the pump field is\cite{Mukhanovetal}
$z_{\rm SSFI} \equiv a \sqrt{2 \epsilon_1}= z\left(1-2\epsilon_1/3
\right)^{1/2}$,
differing from $z$ by a first order term in $\epsilon_1$.
It follows that $U = U_{{\rm SSFI}} + a^2 H^2
\epsilon_1 \epsilon_2 + {\rm order} (\epsilon_i^3)$,
where we have used that  $dz/d\tau \simeq zaH$
at lowest order.
Thus, the second order correction $ \propto \epsilon_1
\epsilon_2$ allows us to compute $U$ up to second order in the
slow-roll expansion from $U_{\rm SSFI}$ given in\cite{Dominique}.

Eq.~(\ref{eq:canon}) can be solved in terms of Hankel functions.
After normalization to vacuum fluctuations for $\sqrt{-w} k/(aH) \to \infty$,
and on expanding the solution in the slow-roll parameters we find
\be
P^{1/2}_{S}(k) = \sqrt{\frac{k^2}{2 \pi^2}} \left| \frac{v_\bk}{z M_{\rm Pl}}
\right|=  \frac{1 -  ( C + 1 - \alpha) \epsilon_1 - {1}/{2} C \epsilon_2}
{2 \sqrt{2} \pi}  \left. \frac{ H }{ M_{\rm Pl}
\sqrt{\epsilon_1}} \right|_{k=aH}, \label{eq:scalarspec}
\ee
where $C$ is a numerical
constant.
At this order in the slow-roll expansion, 
the parameter $\alpha$ is the only difference between SSFI inflation 
and tachyon inflation. It vanishes
in the first case but takes the value $\alpha=1/6$ in the second.
The spectrum of gravity waves $P_T$ in tachyon inflation is exactly as
in SSFI since in absence of anisotropic stress
gravity waves are decoupled from matter. 

A non-vanishing $\alpha$ is responsible for second order deviations
in one of the consistency relations, 
conditions on the observable parameters,
thought to be distinctive of SSFI\cite{Liddleetal}.
Indeed, as expected, to lowest order in the
slow-roll parameters $r=P_T/P_S$, 
$n = 1+ {d \ln {P}_{S}(k)}/{d \ln k}$, and 
$n_T = {d \ln {P}_T(k)}/{d \ln k}$ are identical to those of SSFI,
and the consistency relations are the same.
However at higher order we find 
\be
n_T =-(r/8) \left[1 - (1 - 2 \alpha)r/{16}
 + (1-n) \right]. \label{eq:c2}
\ee
This consistency relation is the next order version of 
$n_T=-2 \epsilon_1 = -r/8$.
There is a deviation from SSFI, represented by a nonvanishing $\alpha$,
which
could in principle be a way of distinguishing
tachyon inflation from SSFI.
However,
in order to see this deviation, $n$,
$r$, and $n_T$ should be known to a precision of $\sim
10^{-3}r^2$.  The error on $n_T$ for future Cosmic
Microwave Background observations
has been  estimated by Song and Knox\cite{Knox}. Even
for the largest possible values
of $r$, $r \sim 1 $,
it is too large
for the deviations predicted by the tachyon
to be observable, so that Eq.~(\ref{eq:c2})
will be very difficult to test.

\section{Models and comparison to WMAP data}

\begin{figure}
\begin{center}
\includegraphics[width=4in]{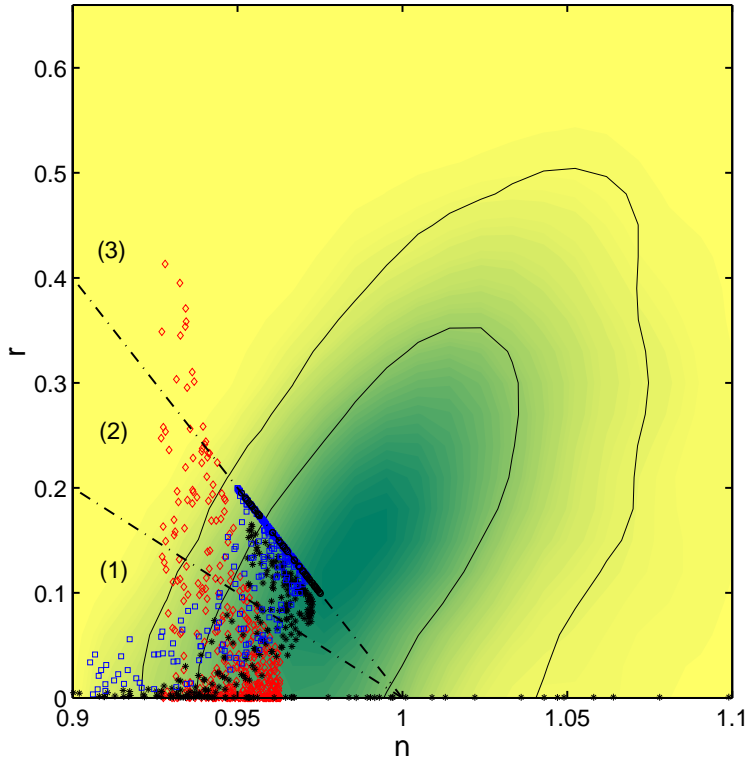}
\end{center}
\caption{Models of tachyon inflation compared to the 2-dimensional
likelihood contours (at $1\sigma$ and $2\sigma$) on the
$(n,r)$-plane.  The points represent the result of a random
sampling from three different tachyonic potentials. In the text: $V_a$, 
squares, blue; $V_b$, circles,
black; $V_c$, stars, black;
and $V_d$, diamonds, red. We consider
$40 \leq N_* \leq 70$. The two dashed
lines correspond to the limits between the three different regimes
of inflation.} \label{fig:sam}
\end{figure}

We now study tachyon inflation for different potentials $V(T)$ and
extract $n$, $r$, and $d n/d \ln k$. We follow the standard
procedure: 1) for a given potential compute
$\epsilon_{1}$, $\epsilon_{2}$, and $\epsilon_2 \epsilon_{3}$ as a
function of $T$; 2) estimate $T_e$, the value of $T$ at the end of
inflation when $\epsilon_1(T_e)=1$; 3) calculate the number of
$e$-foldings as a function of the field $T$; 4) from
$\epsilon_{1}$, $\epsilon_{2}$, and $\epsilon_{3}$ calculate the
observable parameters as a function of $T$ and evaluate them at
$T_*=T(N_*)$, the value of
$T$ at which a length scale crosses the Hubble radius during
inflation. By doing this we can draw some general properties of
tachyon inflation and compare these with WMAP data. 
Results are shown in
Fig.~\ref{fig:sam} using the likelihood contour from the analysis of
S.~Leach and A.~Liddle \cite{Sam}.

We consider three class of potentials, inverse
$\cosh$ potentials, $V_a(T) = \lambda/\cosh(T/T_0)$, the exponential 
potential $V_b(T)=\lambda e^{-T/T_0}$, the potential 
$V_c(T)=\lambda[1+(T/T_0)^4/27]e^{-T/T_0}$ and the inverse power-law 
$V_d={\lambda}/{ [1+(T/T_0)^4]}$. 
The first two are often 
considered in the 
string theory literature\cite{bsft,sen2} 
while the last two are not particulary string 
motivated but display interesting properties.
It is useful to define the constant dimensionless ratio\cite{Malcolm}
$X_0^2 = { \lambda T_0^2}/{M_{\rm Pl}^2} $
which appears in the slow-roll parameters derived from these potentials.
Typically $X_0 \gg 1$ in order for the slow-roll conditions to
be satisfied. 

These
potentials  generally have a
red spectrum of scalar perturbations with a negative and very
small running of the scalar spectral index. For specific choices
of potential such as $V_c$, blue spectra
can be obtained with very small $r$. We divide the $(n,r)$-plane
into three regions of interest corresponding to
1) $V'' \leq 0$, $6 \epsilon_1 \leq \epsilon_2$;
2) $0 < V'' < V'{}^2/ V$, $2 \epsilon_1 < \epsilon_2 <
6 \epsilon_1 $; and 3) $V'{}^2/ V \leq V''  $, $\epsilon_2
\leq 2 \epsilon_1 $.
It is interesting to compare
the predictions of our potentials with the current data and to see 
whether it is
possible to discriminate between SSFI and tachyon inflation.
For $V_a$ inflation takes
place in both regimes 1 and 2. For a large set of parameters $N_*$
and $X_0$ (excluding very small $X_0$), the predictions are well
inside the $2\sigma$ contour. There are non negligible gravity
waves for large $X_0$, though for the range of $N_*$ given above,
$r \lsim 0.2$. When $X_0 \to \infty$, the predictions concentrate
on the line $\epsilon_2 = 2 \epsilon_1$ which are just those of
potential $V_b$. Potential $V_c$ can occupy regimes 1,
2, and 3, and leads to a large contribution of gravity waves,
although $r \lsim 0.2$ in the region not excluded by current data.
Potential $V_d$
occupies much of the region of the inverse cosh potential as well
as yielding blue spectra for negligible $r$.

All the presented models seem to be consistent with the data.
Hence, the first-year WMAP results
are still too crude to constraint significantly the region of
parameters. On the other hand, we still lack of information about
the mechanism of reheating that could take place after tachyon
inflation, leaving us with a large uncertainty on $N_*$. Progress
can be made by better estimating this particular parameter.

Our results point to the fact that it is difficult to distinguish
between a model where inflation is driven by a Klein-Gordon scalar
field or by some other field satisfying a non standard action.
However, none of the potentials we have considered in our
analysis
lead to
both a blue scalar spectral index and large gravity wave spectrum.
Therefore for these potentials a large region in the $(n,r)$-plane
is not probed by tachyon inflation. 
This corresponds, in SSFI, to
the region occupied by hybrid inflation. Detection of $n>1$ and
large $r$, or of a large running of $n$, can lead to the exclusion
of tachyonic inflation.


\begin{thebibliography}{100}
\bibitem{WMAP1} C.~L.~Bennett et al., Astrophys. J. Suppl. {\bf 148} (2003) 1.


\bibitem{Gibbons1}
G.~Gibbons, 
Phys. Lett. B {\bf 537} (2002) 1-4.

\bibitem{action}
A.~Sen, JHEP {\bf 9910} (1999) 008.

\bibitem{DF}
D.~Steer, F.~Vernizzi, accepted for Phys.\ Rev.\ D, [arXiv:hep-th/0310139].

\bibitem{Dominique}
D.~J.~Schwarz, C.~A.~Terrero-Escalante, A.~A.~Garcia,
Phys.\ Lett.\ B {\bf 517} (2001) 243.

\bibitem{GarriMuka}
J.~Garriga, V.~F.~Mukhanov, Phys.\ Lett.\ B {\bf 458}
(1999) 219-225.

\bibitem{Mukhanovetal}
V.~F.~Mukhanov, H.~A.~Feldman, and R.~H.~Brandenberger, Phys.\ Rep.\
{\bf 215}  (1992 ) 203-333.

\bibitem{Liddleetal}
J.~E.~Lidsey, A.~R.~Liddle, E.~W.~Kolb, E.~J.~Copeland,
T.~Barreiro and M.~Abney,
Rev.\ Mod.\ Phys.\ {\bf 69} (1997) 373.

\bibitem{Knox}
Y.~S.~Song and L.~Knox, Phys.\ Rev.\ D {\bf 68} (2003) 043518.

\bibitem{Sam}
S.~M.~Leach and A.~R.~Liddle, Phys.\ Rev.\ D {\bf 68} (2003) 123508.

\bibitem{bsft}
N.~Lambert, H.~Liu and J.~Maldacena,
``Closed strings from decaying D-branes,''
[arXiv:hep-th/0303139].


\bibitem{sen2}
A.~Sen,
JHEP {\bf 9809} (1998) 023;
JHEP {\bf 9812} (1998) 021.

\bibitem{Malcolm}
M.~Fairbairn and M.~H.~Tytgat,
Phys.\ Lett.\ B {\bf 546} (2002) 1.


\end{thebibliography}
\end{document}